# Resonance Energy Transfer and Quantum Entanglement Mediated by Epsilon-Near-Zero and Other Plasmonic Waveguide Systems

Ying Li,[1] Andrei Nemilentsau,[2] and Christos Argyropoulos[1,*]

[1]Dept. of Electrical & Computer Engineering, University of Nebraska-Lincoln, Lincoln, NE, 68588, USA

[2]Dept. of Electrical & Computer Engineering, University of Minnesota, Minneapolis, MN, 55455, USA

*christos.argyropoulos@unl.edu

*The entanglement and resonance energy transfer between two-level quantum emitters are typically limited to sub-wavelength distances due to the inherently short-range nature of the dipole-dipole interactions. Moreover, the entanglement of quantum systems is hard to preserve for a long time period due to decoherence and dephasing mainly caused by radiative and nonradiative losses. In this work, we outperform the aforementioned limitations by presenting efficient long-range inter-emitter entanglement and large enhancement of resonance energy transfer between two optical qubits mediated by epsilon-near-zero (ENZ) and other plasmonic waveguide types, such as V-shaped grooves and cylindrical nanorods. More importantly, we explicitly demonstrate that the ENZ waveguide resonant energy transfer and entanglement performance drastically outperforms the other waveguide systems. Only the excited ENZ mode has an infinite phase velocity combined with a strong and homogeneous electric field distribution, which leads to a giant energy transfer and efficient entanglement independent to the*

*emitters' separation distances and nanoscale positions in the ENZ nanowaveguide, an advantageous feature that can potentially accommodate multi-qubit entanglement. Moreover, the transient entanglement can be further improved and become almost independent of the detrimental decoherence effect when an optically active (gain) medium is embedded inside the ENZ waveguide. We also present that efficient steady-state entanglement can be achieved by using a coherent external pumping scheme. Finally, we report a practical way to detect the steady-state entanglement by computing the second-order correlation function. The presented findings stress the importance of plasmonic ENZ waveguides in the design of the envisioned on-chip quantum communication and information processing plasmonic nanodevices.*

**1 Introduction**

One of the main limitations of the current quantum photonic systems is the rapid loss of spatial and temporal coherence [1,2]. For instance, Förster resonance energy transfer [3], a well-known dipole-dipole interaction between quantum emitters important to light sources, biomedical imaging, and photovoltaic applications, is limited to subwavelength ranges [4,5]. In addition, quantum entanglement [6], which is significant for a variety of emerging applications in quantum communication and computing [7], usually takes place at extremely short distances and for very short time periods, due to the decoherence associated with unavoidable coupling between the system and the surrounding environment [8].

During the last years, considerable research efforts have been dedicated to significantly improve coherence based on the emerging field of quantum plasmonic metamaterials [9,10]. These artificially engineered nanostructures can serve as a novel platform to trigger, harness, and enhance coherent light-matter interactions at the nanoscale [9,11]. For instance, long-range energy transfer [12–15] and quantum entanglement [16–19] can be achieved by coupling quantum emitters to the plasmonic modes propagating along commonly used plasmonic waveguide configurations, such as V-shaped grooves milled in a flat metallic surface and cylindrical nanorods. Moreover, successful subwavelength waveguiding [20] and single plasmon generation [21] were experimentally demonstrated based on these systems. However, the emitters' positions in these plasmonic nanowaveguides are crucial to the efficiency of the aforementioned quantum processes leading to rapidly varying energy transfer and entanglement as a function of the emitters' separation distance. Therefore, creating and preserving over extended time periods and long distances the efficient entanglement and strong energy transfer between quantum dipole emitters randomly distributed in a nanophotonic system remains a key challenge.

Metamaterials exhibiting epsilon-near-zero (ENZ) behavior promise to overcome this limitation [22]. They have attracted increased attention due to the large field enhancement in combination with a uniform distribution, leading to various applications in enhancing fluorescence, squeezing and tunneling light through bended waveguides, coherent perfect absorption, and boosting optical nonlinear effects [22–24]. Recently, some interesting works have been proposed that connect ENZ metamaterials, even zero index metamaterials (ZIM), with qubit entanglement [25,26]. In these configurations, the effective near zero permittivity has been designed by utilizing bulk materials operating at their plasma frequencies, complex layered hyperbolic metamaterial structures [26] or all-dielectric photonic crystal waveguides [25]. However, all of these approaches to achieve an ENZ response suffer from fabrication limitations, limited tunability, and very narrowband responses, which make them impractical for realistic applications. In addition, and more importantly, the field enhancement within these ENZ materials is relatively low and has limited spatial extent, leading to relatively weak light-matter interactions.

In this work, we explicitly prove that a specific ENZ plasmonic configuration can substantially improve the quantum entanglement and the resonance energy transfer compared to other commonly used plasmonic waveguide configurations, such as V-shaped grooves milled in a flat metallic surface and cylindrical nanorods. The proposed quantum ENZ metamaterial system can support efficient long-range resonance energy transfer (RET) and entanglement between quantum dipole emitters independent of their positions within the nanowaveguide, over extended time periods and long separation distances. It is comprised of an array of metallic waveguides that exhibit an effective ENZ response at their cutoff frequency in combination with enhanced and homogeneous electromagnetic fields inside their nanochannels [27–32]. These interesting features, combined with the strong omnidirectional resonant coupling at the ENZ frequency, are ideal conditions to boost coherent light-matter interactions along elongated regions and can increase the temporal and spatial coherence between different emitters leading to multi-qubit entanglement [33]. For example, it was recently shown that long-range superradiance can be sustained by using this system [30].

Here, we demonstrate quantum entanglement that persists over extended time periods and long distances due to the ENZ mode formed inside the plasmonic nanowaveguide channels. More importantly, it is demonstrated that the introduction of active (gain) media inside the plasmonic ENZ nanochannels can lead to perfect loss compensation of the inherent plasmonic losses which, subsequently, further boosts transient quantum entanglement, making this effect almost independent to the detrimental influence of decoherence. Enhanced RET is also presented based on the proposed ENZ plasmonic system for large donor-acceptor separation distances and independent of their positions. Finally, efficient steady-state entanglement, independent of the emitters' distances and positions, is presented and a practical way to detect it by computing the second-order correlation function is proposed. The findings of this work stress the importance of plasmonic ENZ waveguides as an ideal nanophotonic platform for supporting efficient quantum electrodynamic effects.

## 2 Enhanced Resonance Energy Transfer

We investigate the enhancement of the RET effect between a donor (D) and acceptor (A) pair mediated by the proposed ENZ and the other plasmonic waveguide types. We demonstrate giant RET for the ENZ case independent of the D-A pair distance and enhanced RET, but oscillating with the D-A pair distance, for the other plasmonic waveguide cases. The unit cell geometry of the proposed ENZ plasmonic waveguide is illustrated in Fig. 1(a). It is composed of a rectangular slit carved in a silver screen and loaded with a dielectric material. The slits (cyan) are loaded with glass with a relative permittivity equal to $\varepsilon = 2.2$, and the silver permittivity dispersion values are taken from previously obtained experimental data [34]. This free-standing waveguide geometry was originally introduced in [27] and can sustain an ENZ resonance at the cutoff wavelength of its dominant quasi-TE$_{10}$ mode. The extraordinary optical transmission combined with large field enhancement and uniform phase distribution inside the nanoslits was demonstrated at the ENZ operation wavelength [30,31]. This is due to an anomalous impedance matching phenomenon that depends only on the interface properties, i.e., on the aperture to period ratio of the array, and is therefore independent of the grating periodicity and the waveguide channel thickness [30,31]. Here, the slit dimensions have width $w = 200\,\text{nm}$, height $t = 40 \text{nm}$ ($t \ll w$), and length $l = 1 \mu\text{m}$, respectively, and the grating period is equal to $a = b = 400 \text{nm}$. To clearly demonstrate the aforementioned ENZ response, we operate close to the cutoff wavelength of the proposed waveguide, which is $l$ = 1016 nm. At this frequency, the effective electromagnetic properties of the waveguides' plasmonic channels become equivalent to the properties of a bulk ENZ metamaterial exhibiting uniformly enhanced electric field distribution, as illustrated in the right inset of Fig. 2, plotted along the xz-plane of the unit cell geometry shown in Fig. 1(a) [30,31].

Next, we consider a pair of identical two-level quantum emitters as the D-A pair located at the center of the plasmonic waveguide channel along the x-axis [see Fig. 1(a)]. We assume that the emitter A with dipole moment $\mu_A$ is located at the arbitrary position $\mathbf{r}_A$ and the emitter D with identical dipole moment $\mathbf{\mu_D} = \mathbf{\mu_A}$ is placed at the position $\mathbf{r}_D$. In order to quantify the RET, we use the following normalized energy transfer rate (nETR) formula: [5,12]

$$\text{nETR} = \frac{\text{Im}\left[\mathbf{\mu}_A^* \cdot \mathbf{G}(\mathbf{r}_A,\mathbf{r}_D) \cdot \mathbf{\mu_D}\right]}{\text{Im}\left[\mathbf{\mu}_A^* \cdot \mathbf{G}_{\text{vac}}(\mathbf{r}_A,\mathbf{r}_D) \cdot \mathbf{\mu_D}\right]} = \frac{\left|\mathbf{n}_A \cdot \mathbf{E}_D(\mathbf{r}_A)\right|^2}{\left|\mathbf{n}_A \cdot \mathbf{E}_{D,\text{vac}}(\mathbf{r}_A)\right|^2} \tag{1}$$

where $\mathbf{\mu}_A^* = \mathbf{\mu}_D^*$ is the complex conjugate of the transition dipole moment of the emitter and $\mathbf{n}$ is the unit vector in the direction of the dipole moment ($\mathbf{\mu_A} = \mathbf{\mu_D} = \mu \cdot \mathbf{n}$). $\mathbf{G}(\mathbf{r}_A,\mathbf{r}_D)$ [$\mathbf{G}_{\text{vac}}(\mathbf{r}_A,\mathbf{r}_D)$] is the system's dyadic Green's function in the ENZ waveguide (vacuum), while $\mathbf{E}_D(\mathbf{r}_A)$ ($\mathbf{E}_{D,\text{vac}}(\mathbf{r}_A)$) is the electric field induced at the acceptor location $\mathbf{r}_A$ by an electric dipole at the source (donor) point $\mathbf{r}_D$ when both emitters are inside the waveguide (in the vacuum). Therefore, the nETR represents the energy transfer enhancement in the presence of the waveguide system with respect to vacuum.

We also introduce two alternative designs of finite plasmonic waveguides, a V-shaped channel cut in a metal plane and a cylindrical metallic nanowire, as shown in Figs. 1(b) and 1(c), and compare their performance with the aforementioned ENZ nanowaveguide. In the former case [Fig. 1(b)], we choose the depth of the groove to be equal to $L = 235$ nm and the opening angle to be $\theta = 10°$; in the nanowire configuration [Fig. 1(c)], the radius is fixed to R = 25 nm. The lower panels (d)-(f) in Fig. 1 display the electric field spatial distribution of the supported fundamental resonant mode for each of the plasmonic waveguides. We want to point out that all the plasmonic waveguides studied in our work have finite length, which makes a significant difference compared to the infinite-length plasmonic waveguide designs presented in [12]. The finite length leads to stronger emitter-plasmon interactions and a more fair performance comparison between the different plasmonic waveguide systems [18]. The channel lengths are chosen to be $l = 1\mu m$ for the ENZ and cylindrical rod (nanowire) plasmonic waveguides and $l = 1.4\mu m$ for the groove plasmonic waveguide. Unlike the cases of the ENZ and the rod waveguides, where the mode electric field is strong at the waveguides ends, the modal field is weak at the groove ends. Thus we choose the groove's channel length to be longer in order to ensure that the distance between two antinodes located near the groove ends is also $1\mu m$. We place the emitters in the area of the strong electric fields in order to achieve a fair performance comparison between the different plasmonic waveguides. Particularly, the D-A pair is embedded in the center of the dielectric slit inside the ENZ waveguide. In the case of the two other waveguides, the emitters are placed at a height equal to $h = 20\,nm$ above them.

We always assume that the dipole moments of the D-A pair are identical and polarized vertically (along the z-direction) in the cases of ENZ and rod waveguides and horizontally (along the y-axis) in the groove waveguide in order to achieve optimal coupling with their electromagnetic resonant modes [17]. The dipole transition wavelength of the emitters is 1018 nm, which is near the ENZ cutoff resonant wavelength ($\lambda = 1016 \text{nm}$). The geometric parameters of the finite groove and rod waveguides are chosen with the goal to have the same propagation length (about 17 um) at the operating wavelength $\lambda = 1018 \text{nm}$. In our case, the propagation length of the ENZ plasmonic waveguide is shorter (approximately 1.7 um) due to the higher plasmonic losses compared to the other two waveguides. However, the exceptional field enhancement properties of the ENZ configuration alleviate this problem and lead to much more efficient energy transfer over longer distances compared to the other two waveguide designs, as it will be demonstrated in the following.

To study the energy transfer rate for each waveguide, we fix the location of emitter A at the edge of each channel for the ENZ and rod waveguides and at the antinode location near the groove end for the groove waveguide case. The emitter D is gradually moved along the x-axis. The nETR is calculated according to Eq. (1) and illustrated in Fig. 2 as a function of the separation distance between the D-A emitter pair. Note that the transfer rate in the case of the ENZ waveguide (black line) gradually grows with increasing separation distances, and can reach a gigantic six orders of magnitude enhancement compared to the free space scenario. Whereas, the transfer rate enhancement for groove and rod waveguides oscillates with the separation distance and cannot exceed four orders of magnitude enhancement. The large enhancement of RET, especially for the ENZ waveguide case, is directly related to the field distribution properties shown in the right inset of Fig. 2, and computed by using full-wave numerical simulations (COMSOL Multiphysics), where an excited donor (D) with a unit electric current dipole moment of $1\,\mathrm{A}\cdot\mathrm{m}$ is located in the center or slightly to the right of the center (rod) of each waveguide, where the maximum field enhancement of each resonant mode occurs. It is clear that the ENZ plasmonic waveguide shows much larger field enhancement combined with uniform field distribution along the entire channel in contrast to the other two waveguide types, where a standing wave distribution is observed along the finite-length channels due to Fabry-Perot (FP) interference. Therefore, a substantial two orders of magnitude RET enhancement is achieved by using the ENZ waveguide compared to the other two waveguide types, which, very importantly, is also independent to the emitters' separation distance. It is also worth to point out that, unlike the infinite waveguide case [12], the nETR does not continuously increase in the finite groove or rod waveguides. The nETR rapidly oscillates as a function of the separation distance because the plasmon reflections from the finite waveguides' ends can increase or decrease qubit-waveguide coupling due to constructive or destructive interference [18]. Note that the large increase of RET rates shown in Fig. 2 with an increment of the distance between the D-A emitter pair is counterintuitive, since ohmic losses exist in all three plasmonic waveguide cases. This is because the used channel lengths are always smaller than the propagation lengths for all currently investigated cases. Hence, the absorption of each waveguide propagating mode is limited and does not cause the RET to decrease for the currently used separation distances [12]. To sum up, the most important result of this section is the gigantic long-range RET enhancement, substantially surpassing the near-field region, obtained only for the ENZ waveguide (Fig. 2), which is also independent to the D-A emitter pair separation distance. Such a gigantic long-range RET enhancement has not been reported before [35–39], making ENZ waveguides an ideal nanophotonic environment to control and enhance RET-based applications.

## 3 Robust Entanglement Mediated by Plasmonic Waveguides

In this section, we present long-range quantum entanglement between a pair of two-level quantum emitters mediated by the ENZ plasmonic waveguide, which persists over extended time periods and long distances. The response is compared to the other two plasmonic waveguides presented in Figs. 1(b), (c) and the benefits of the ENZ case are stressed. The theory of both transient and steady-state quantum entanglement, quantified by computing the concurrence metric, is briefly introduced. Then, we calculate the entanglement and present ways to further improve it by the incorporation of active media inside the ENZ waveguides. Finally, we describe how to achieve the steady-state enhanced entanglement and propose a practical way to detect it by computing the second-order correlation function. In all these results, the ENZ waveguide system demonstrates a substantially improved quantum optical performance compared to the other plasmonic waveguide systems.

### 3.1. Entanglement Theory

We consider two identical two-level emitters (also known as qubits) with the same emission frequency $\omega_0$ placed inside different plasmonic waveguides, as schematically illustrated in Figs. 1(a)-(c). The dynamic evolution of quantum systems coupled to lossy plasmonic environments can be fully characterized by the dyadic Green's function in combination with the formalism of master equations [40,41]. The former is a classical quantity that is widely used to study the spontaneous decay of quantum emitters coupled to lossless or lossy structures [5]. The latter is a quantum equation that is used to describe the dynamics of the density matrix $\rho$ of the two-qubit system in the vicinity of a plasmonic reservoir [4]. Assuming weak excitation (no saturation) and operation in the weak coupling regime, the Born-Markov and rotating wave approximations can be applied to compute the master equation given by: [40,41]

$$\frac{\partial \rho}{\partial t} = \frac{1}{i\hbar}[H,\rho] - \frac{1}{2}\sum_{i,j=1}^{2}\gamma_{ij}\left(\rho\sigma_i^\dagger\sigma_j + \sigma_i^\dagger\sigma_j\rho - 2\sigma_i\rho\sigma_j^\dagger\right), \qquad (2)$$

where the Hamiltonian characterizing the coherent part of the dynamics in Eq. (2) is equal to:

$$H = \sum_i \hbar(\omega_0 + g_{ii})\sigma_i^\dagger\sigma_i + \sum_{i\neq j}\hbar g_{ij}\sigma_i^\dagger\sigma_j. \qquad (3)$$

In the above equations, $\rho$ is the density matrix of the system of two identical qubits; and $\sigma_i$ ($\sigma_i^\dagger$) is the destruction (creation) operator applied to the $i$-th qubit. The Lamb shift $g_{ii}$ is due to the self-interaction of each qubit placed inside the ENZ waveguide and is usually much lower (GHz range) compared to the higher THz optical frequencies involved in our study [42]. Thus, it can be neglected, especially for distances between the emitter and the metallic walls larger than 10 nm [5]. The other part of Eq. (3), given by $g_{ii}$, characterizes the coherent dipole-dipole interactions and can be computed by the formula: [17]

$$g_{ij} = \left(\omega_0^2/\varepsilon_0\hbar c^2\right)\mathrm{Re}[\boldsymbol{\mu}_i^* \cdot \mathbf{G}(\mathbf{r}_i,\mathbf{r}_j,\omega_0)\cdot\boldsymbol{\mu}_j], \qquad (4)$$

where the Green's tensor $\mathbf{G}(\mathbf{r}_i,\mathbf{r}_j,\omega_0)$ satisfies the classical Maxwell equations for a point dipole source located at an arbitrary spatial position $\mathbf{r}_j$ [5]. Equation (4) characterizes the dipole-dipole coupling between the qubits placed in spatial points $\mathbf{r}_j$ and $\mathbf{r}_i$. The parameters $\gamma_{ij}$ in the dissipative and noncoherent term of the master equation (2) are given as a function of the imaginary part of the Green's tensor, which is also referred to as local density of states (LDOS): [17]

$$\gamma_{ij} = \left(2\omega_0^2 / \varepsilon_0 \hbar c^2\right) \text{Im}[\boldsymbol{\mu}_i^* \cdot \mathbf{G}(\mathbf{r}_i, \mathbf{r}_j, \omega_0) \cdot \boldsymbol{\mu}_j]. \tag{5}$$

Equation (5) can be used to compute the decay rate induced by self-interactions $\left(\gamma_{ii}\right)$, also known as spontaneous emission rate, and mutual interactions $\left(\gamma_{ij}\right)$. Therefore, $\gamma_{12}$ represents the contribution to the decay rate of Emitter 1 at an arbitrary position $\mathbf{r}_1$ due to interference caused by Emitter 2 located at an arbitrary position $\mathbf{r}_2$ inside the plasmonic waveguides. Both Eqs. (4) and (5) can be computed by solving Maxwell's equations, either analytically or numerically, through full-wave simulations. Note that the relations $\gamma_{ij} = \gamma_{ji}$ and $g_{ij} = g_{ji}$ are always valid in the currently studied reciprocal waveguides because of the Green's dyadic symmetry.

The goal is to solve Eq. (2) for the proposed ENZ and the other plasmonic waveguide systems. In order to solve this equation and compute the density matrix, a convenient basis for the vector space of the two-qubit system needs to be defined. As we study identical emitters placed into equivalent positions, i.e. $\gamma_{11} = \gamma_{22} = \gamma$, it is convenient to work in the Dicke basis: $|3\rangle = |e_1\rangle \otimes |e_2\rangle = |e_1, e_2\rangle$, $|0\rangle = |g_1\rangle \otimes |g_2\rangle = |g_1, g_2\rangle$, and $|\pm\rangle = 1/\sqrt{2}\left(|e_1, g_2\rangle \pm |g_1, e_2\rangle\right)$, where $|e_i\rangle$ is the excited state of the i-th qubit, while $|g_i\rangle$ is the ground state of the i-th qubit. This basis is the appropriate way to characterize the response of two-qubit systems since it leads to a diagonalized Hamiltonian [Eq. (3)] [16–18]. A schematic of the collective states of two identical emitters coupled to a dissipative plasmonic reservoir is given in Fig. 3, where it can be seen that the ground state $|0\rangle$ and the excited state $|3\rangle$ are not affected by the dipole-dipole interactions. However, the dipole-dipole interactions lead to a shift at the energies of the symmetric $|+\rangle$ and antisymmetric $|-\rangle$ collective states by $\pm g_{12}$ compared to their energies in the case when the dipole-dipole interactions are negligible. Moreover, the qubit-qubit dissipative coupling induces the modified collective decay rates $\gamma + \gamma_{12}$ and $\gamma - \gamma_{12}$, which correspond to superradiant and subradiant states, respectively [41]. It is shown in the next section that in the case of the ENZ waveguide $\gamma = \gamma_{12}$ and we observe a pure superradiant emission and a zero subradiant decay rate independent to the emitters' separation distance [30,43]. This is different compared to the collective response of the groove and rod plasmonic waveguides.

The master equation (2) is solved in the two qubits state basis shown in Fig. 3. Only one of the qubits is assumed to be excited initially, thus $\rho_{33}(t) = 0$ and $\rho_{00}(0) = 0$, since exciting two qubits in the same time is a very ineffective way to produce entanglement [44]. Hence, the system is prepared in the unentangled state $|e_1, g_2\rangle = 1/\sqrt{2}(|+\rangle + |-\rangle)$. In this case, the only nonzero components of the density matrix at $t = 0$ are $\rho_{++}(0) = \rho_{+-}(0) = \rho_{-+}(0) = \rho_{--}(0) = 1/2$; and the density matrix elements simply become:

$$\rho_{++}(t) = 0.5 e^{-(\gamma + \gamma_{12})t} \tag{6}$$

$$\rho_{--}(t) = 0.5 e^{-(\gamma - \gamma_{12})t} \tag{7}$$

$$\rho_{+-}(t) = 0.5 e^{-(\gamma - 2ig_{12})t} \tag{8}$$

$$\rho_{-+}(t) = 0.5 e^{-(\gamma + 2ig_{12})t}. \tag{9}$$

In general, the entanglement between two emitters can be quantified by computing the concurrence C that was originally introduced by Wootters [45]. This quantity is defined as: $C = \max(0, \sqrt{u_1} - \sqrt{u_2} - \sqrt{u_3} - \sqrt{u_4})$, where $u_i$ are the eigenvalues of the matrix $\rho\tilde{\rho}$ and $\tilde{\rho} = \sigma_y \otimes \sigma_y \rho^* \sigma_y \otimes \sigma_y$ is the spin-flip density matrix, with $\sigma_y$ being the Pauli matrix. The value of concurrence can vary between zero (unentangled state) to one (completely entangled qubits). If the density matrix of the system is characterized by Eqs. (6)-(9), where only one emitter is initially in the excited state, then the concurrence can be simplified and computed by the formula: [18]

$$C(t) = \sqrt{(\rho_{++} - \rho_{--})^2 + 4\operatorname{Im}(\rho_{+-})^2}. \tag{10}$$

After substituting Eqs. (6)-(9) in Eq. (10), it takes the final form:

$$C(t) = 0.5\sqrt{\left[e^{-(\gamma+\gamma_{12})t} - e^{-(\gamma-\gamma_{12})t}\right]^2 + 4e^{-2\gamma t}\sin^2(2g_{12}t)}. \tag{11}$$

The derived transient concurrence [Eq.(11)] is a very useful quantity that can provide physical insights about the entanglement process between two emitters when only one of the emitters is excited. First, it can be seen that $C(0) = 0$ which is expected because the system is initially at an unentangled state. As the time progresses, $t > 0$, the concurrence becomes larger than zero meaning that the emitters become entangled. However, at some point the concurrence starts to decay with time and becomes zero again, $C(t) = 0$, after a long period of time $(t \to \infty)$. In addition, the concurrence can strongly oscillate for short time durations $(t < 1/2\gamma)$ if the coherent dipole-dipole interactions are large $(g_{12} \gg \gamma)$. This is usually the case for a photonic-crystal or a microcavity-based entanglement [46,47] that can lead to high concurrence values; but for a very short time duration. Notice that for the ideal case of a lossless and infinite waveguide, $\gamma = \gamma_{12}$, i.e., the decay rate of the asymmetric state $|-\rangle$ is zero, so the entanglement can grow with time monotonically and obtain a steady-state with concurrence up to the maximum value of $C = 0.5$ [16,17,48]. It is demonstrated in the next section that the proposed ENZ waveguide system satisfies the ideal conditions, i.e., $g_{12} \ll \gamma$ and $\gamma = \gamma_{12}$, and achieves long lasting strong concurrence without the onset of oscillations, especially when active media are incorporated.

In order to prevent the transient concurrence from decaying after some time and obtain a more practical steady-state entanglement, external pumps with the same frequency ($\omega_p$) can be used to individually pump each emitter or an emitter cluster loaded in the plasmonic waveguides. In this case, an additional term $1/i\hbar[V,\rho]$ needs to be introduced in the right hand side of the master equation (2), where the operator:

$$V = -\sum_{i}^{2} \hbar \left( \Omega_i e^{-i\Delta_i t} \sigma_i^\dagger + \Omega_i^* e^{i\Delta_i t} \sigma_i \right), \tag{12}$$

characterizes the interaction between the pump field and the emitter. [18] The parameter $\Omega_i = \mu \cdot E_{0i}/\hbar$ is the effective Rabi frequency of the pump that depends on the field $E_{0i}$ induced at the pumped $i$-th emitter inside the ENZ or the other waveguides. The parameter $\Delta_i = \omega_0 - \omega_p$ is the detuning parameter due to the pump frequency $\omega_p$. After expressing $\rho$ in the usual basis $|e_1,e_2\rangle$, $|e_1,g_2\rangle$, $|g_1,e_2\rangle$, and $|g_1,g_2\rangle$, we can calculate the steady state concurrence by solving numerically the modified versions of the previously derived differential equations with density matrix solutions given by Eqs. (6)-(9), where we have included the Rabi frequency and detuning parameter due to the external pumping [18].

### 3.2 Results and discussion

### 3.2.1 Passive transient entanglement

The theory presented in the previous section is applied to the ENZ, groove, and rod plasmonic waveguide systems. The emission wavelength of the emitters is $\lambda = 1018$ nm for all cases, almost equal to the cutoff wavelength of the ENZ plasmonic waveguide, where the field is strong and homogeneous along the nanochannel (Figs. 1(d) and 2). Note that the cutoff wavelength of this robust ENZ plasmonic system can be tuned to any value, if the width $w$ of the waveguide's channel is varied [31]. This useful property can be used to accommodate a plethora of different emitters embedded in the proposed ENZ effective medium with various emission frequencies, such as quantum dots [49–51]. In the current case, the two identical qubits are either embedded in the middle of the ENZ waveguide, or placed along the narrow plasmonic channels of the rod and groove waveguides (Figs. 1(a)-(c)). The qubits are always placed at two equivalent (symmetric) positions along the waveguides to ensure that $\gamma_{11} = \gamma_{22} = \gamma$ and separated by a distance $d$. Full-wave simulations are performed to compute the Green's function and, as a consequence, the dipole-dipole interaction coefficients $g_{ij}$ and decay rates $\gamma_{ij}$ given by Eqs. (4) and (5), respectively.

To better understand the collective behavior of the qubits, which is independent of the qubit separation distance only in the case of the ENZ waveguide, we plot in Fig. 4 the computed mutual interaction decay rates $\gamma_{12}$ and dipole-dipole interactions $g_{12}$ as a function of the separation distance between the qubits positioned symmetrically inside the ENZ waveguide. All the results are normalized to $\gamma$, i.e., the computed spontaneous emission decay rate of each emitter. We demonstrated in [30], by using the nonlocal density of optical states, that a strong and homogeneous enhancement of the collective spontaneous emission can be achieved by using the proposed ENZ structure, leading to a perfect superradiance response at the cutoff resonance for large separation distances between the qubits. Interestingly, the results presented in Fig. 4 provide an alternative approach to prove the perfect superradiance decay that does not depend on the qubits separation distance. It can be seen that, as we gradually increase the separation distance of the qubits placed inside the ENZ waveguide channel, the decay ratio $\gamma_{12}/\gamma$ decreases due to the inevitable decoherence induced by the system's radiation and plasmonic losses but the ratio is still close to 1 along the entire channel. This effect is further improved in the case of active media embedded in the ENZ channel, as it will be described in the next section. Therefore, a perfect superradiance decay, independent of the qubits separation distance, is achieved because the self-interactions characterized by $\gamma$ are almost equal to the mutual interactions computed by $\gamma_{12}$ along nearly the entire channel of the ENZ waveguide. As a result, when both qubits are excited, the qubit-qubit dissipative coupling happens along the superradiant collective decay state $|+\rangle$ only (see Fig. 3); and the collective decay rate along the subradiant state $|-\rangle$ is suppressed and becomes almost equal to zero. In addition, we note that the dissipative decay rate ratio $\gamma_{12}/\gamma$ plotted in Fig. 4 is much larger than the normalized coherent coupling term $g_{12}/\gamma$ along the entire ENZ channel, directly indicating that the proposed ENZ system satisfies the ideal condition to achieve the best entanglement performance, i.e., $g_{12} \ll \gamma$ and $\gamma = \gamma_{12}$ [17]. Hence, strong quantum entanglement is expected, that persists over extended time periods and long separation distances, independent of the emitters' positions and distances, by using the proposed ENZ plasmonic waveguide system.

Next, we study the entanglement between the same two qubits placed inside the ENZ waveguide channel by calculating the concurrence given by Eq. (11), where the dissipative and coupling decay rates are shown in Fig. 4. Note that in this case only one qubit is excited, which is different from the superradiance scenario; and the computed transient concurrences can be seen in Fig. 5 (black lines), where the distances between the two qubits are fixed to three values: (a) 100 nm,

(b) 300 nm, and (c) 700 nm, respectively. We also consider the behavior of concurrence for the emitters coupled to several alternative environments, such as a homogeneous dielectric bulk space with relative permittivity $\varepsilon = 2.2$ (green line), and the finite groove (red line) and cylindrical rod (blue line) plasmonic waveguides. It is clear that all the three plasmonic waveguides demonstrate improved and more efficient entanglement compared to the case of a pure dielectric medium, where moderately high concurrence values are only obtained for very small inter-qubit distances (near-field) and for very short time durations (Fig. 5(a)). Strong surface plasmon and ENZ modes are excited along the plasmonic waveguides (Figs. 1(d)-1(f)) that lead to a remarkable confinement and enhancement of the electromagnetic fields in stark contrast to the dielectric medium case. We also clearly observe in Fig. 5 that the transient concurrence, and hence entanglement, mediated by the passive ENZ plasmonic waveguide is superior compared to the other two plasmonic waveguides, even for large emitters' separation distances (Fig. 5(c)). This is due to the homogeneous electromagnetic field mode (inset of Fig. 2) at the cutoff wavelength that spreads across the entire ENZ waveguide geometry, resulting in the large absolute values of the decay rate $\gamma_{12}$ and very small values of the interaction rate $g_{12}$ along the entire dielectric nanochannel even for a passive configuration (Fig. 4). On the contrary, the qubits entanglement mediated by the finite groove and rod waveguides has lower concurrence values and depends strongly on the spatial position of both emitters, which is a severe disadvantage for their practical applications, since it is very difficult to accurately position nanoemitters in the nanoscale. The qubits concurrence in these waveguides depends strongly on the inter-emitter distance due to the standing wave field pattern of the FP cavity mode obtained along each finite-length waveguide channel (inset of Fig. 2). The ENZ waveguide does not suffer from this limitation and the emitters can be located anywhere inside the nanochannel, a property that paves the way to efficient and long-range multi-qubit entanglement persisting over extended time periods and long distances.

### 3.2.2 Active transient entanglement

The qubit entanglement results presented in the previous section are focused on passive structures. Ohmic losses are detrimental to the coherent performance of the proposed waveguides leading to decoherence and, as a result, rapid decrease of entanglement over extended time periods. In this section, we propose an active ENZ plasmonic waveguide design that can compensate or even totally suppress the nonradiative ohmic losses, when active media with extremely small gain coefficients are appropriately loaded inside the ENZ nanochannels. The nonradiative rate is minimized in this case, or even become equal to zero, leading to long-range transient entanglement persisting over more prolonged time durations compared to the passive ENZ configuration.

A schematic of the proposed active ENZ configuration is shown in the right inset of Fig. 6(a), where the geometry is similar to the passive ENZ plasmonic waveguide used in the previous sections (also shown in the left inset of Fig. 6(a)). Now, the waveguide is loaded with an active material (yellow region) with length $l_{ac}$ = 800 nm, slightly smaller than the total channel length $l$ = 1 um, and the rest of the channel is loaded with a lossless dielectric material (i.e., glass, cyan region) similar to the previous section. The permittivity of the active material follows a realistic Lorentz dispersion model obeying Kramers-Kronig relations with relative permittivity: $\varepsilon = \varepsilon_\infty + \varepsilon_{Lorentz}\omega_0^2/(\omega_0^2 - 2i\omega\delta_0 - \omega^2)$ , where $\varepsilon_\infty = 2.196$ , $\varepsilon_{Lorentz} = -0.0247$ , $\omega_0 = 1.433\times10^{15}$ rad/s , $\delta_0 = 1.0\times10^{15}$ rad/s and $\omega = 2\pi f$ [52,53]. Realistic gain materials that exhibit these dispersion values are fluorescent dyes (for example Rhodamine) doped in dielectric materials [54,55]. The active material permittivity values around the ENZ resonance are plotted in the inset of Fig. 6(b). The real part of the permittivity in this case is equal to $\varepsilon_r \approx 2.2$ and the imaginary part (gain) is approximately $\delta \approx 0.012$. It has been proven before [32] that when the gain coefficient of the embedded active material takes this particularly small value (i.e., $\delta \approx 0.012$), the transmission of the ENZ wavelength, when it is illuminated by a plane wave, becomes equal to one, and the inherent plasmonic losses of the waveguide are compensated. This perfect loss compensation mechanism stems from the formation of an exceptional point in the reflection/transmission response of the proposed active ENZ structure, where the eigenvalues of the active system coalesce [32,56]. The exceptional point can be accessed with very low gain values by using the proposed active ENZ waveguide geometry due to the strong and homogeneous field enhancement inside the nanochannel.

The computed transmittance and reflectance of the passive and active ENZ waveguides illuminated by a normally incident plane wave are demonstrated in Fig. 6(a), where the black lines refer to the plasmonic channels loaded with a passive dielectric material (i.e., left inset structure), while the red lines represent the channels loaded with the realistic active material characterized by the aforementioned Lorentzian dispersion model (right inset structure). Note that emitters do not exist in these simulations. In the absence of the gain medium (black lines), resonant optical transmission occurs at the ENZ cut-off wavelength $(\lambda = 1016\,\text{nm})$, as it has been predicted in the past [31,32]. The maximum transmission is small $(T_{\max} \approx 0.3)$ though, since the presented passive ENZ waveguide is a lossy system and inevitable ohmic losses exist due to the metallic (silver) walls. However, the field distribution is still uniform and enhanced inside the nanochannel, even for this lossy scenario, which plays a key role in the enhanced entanglement observation presented in the previous section. When an active material is included in the waveguide channel (red lines), perfect transmittance and zero reflectance are obtained at the ENZ wavelength point, which now coincides with an exceptional point [32]. Perfect loss compensation is achieved at this point, minimizing the nonradiative losses of the active ENZ system and, as a result, leading to reduced decoherence. This property is expected to positively affect the entanglement operation.

Hence, we study the transient entanglement between two similar qubits placed inside the active ENZ waveguide channel and compare its performance with the passive case presented in the previous section. The operating wavelength of the qubits is 1018 nm, which again is near the ENZ cutoff resonant wavelength or exceptional point $(\lambda = 1016\,\text{nm})$, and the qubits' locations are fixed at each waveguides' end having a large, wavelength scale, separation distance on the order of $d = 900\,\text{nm}$. Notice that $d > l_{ac}$, which means that both quantum emitters are inserted in the lossless passive dielectric material to avoid convergence issues of the Green function at the source locations (cyan region in the right inset of Fig. 6(a)). Thus, the concurrence entanglement metric introduced in section 3.1 still works for our active ENZ waveguide system. Interestingly, it can be seen in Fig. 6(b) that the active ENZ plasmonic waveguide (red line) demonstrates stronger and more prolonged transient entanglement compared to the passive case (black line). This is due to the perfect loss compensation achieved by the active ENZ system, where the nonradiative losses are minimized at the exceptional point, leading to reduced decoherence and improved entanglement. The transient concurrence of the active system also decays in time, however with a much slower rate compared to the passive ENZ case, due to radiative losses that are unavoidable in open waveguide systems, similar to the currently studied configurations. A potential solution to this problem is presented in the next section 3.2.3. However, even the currently presented long-range and prolonged time duration transient entanglement achieved by the active ENZ waveguide system can be ideal property to the envisioned on-chip quantum communication and information processing plasmonic nanodevices. Finally, it is worth mentioning that the aforementioned results are still obtained in the weak coupling regime (where the currently used Markov approximation is valid), since the gain coefficient is extremely small and does not strongly influence the scale of the field enhancement inside the nanochannel.

### 3.2.3 Steady-state entanglement and detection

Up to this point, we have clearly demonstrated that the transient entanglement, mediated by the ENZ plasmonic waveguide, persists over long periods of time and long distances and, in addition, is independent of the qubits' positions and separation distances. However, the transient entanglement between the emitters placed inside the ENZ waveguide still disappears after a long time period, even for the active scenario, and becomes equal to zero at steady-state, as it is shown in Figs. 5 and 6(b). This effect is a direct consequence of decoherence caused by the depopulation of the emitters' excited states due to radiation and plasmonic losses. Thus, in order to achieve a high steady-state entanglement with concurrence $C(t \to \infty) \neq 0$, an additional pump excitation applied to the qubits inside the ENZ waveguide is introduced. Figure 7 shows the concurrence of two qubits driven by the external pump in the presence of the proposed passive ENZ plasmonic waveguide compared to the finite groove and rod plasmonic waveguide cases. The emitters' locations are fixed at each waveguide's end with a separation distance $d = 700$nm. In the upper panels of Fig. 7, we use three types of coherent pumps: (a) asymmetric pumping with Rabi frequencies $\Omega_1 \neq 0$, $\Omega_2 = 0$, (b) symmetric pumping with $\Omega_1 = \Omega_2$, and (c) antisymmetric pumping with $\Omega_1 = -\Omega_2$, leading to different relative phases among the pumping lasers applied to each qubit. In the ENZ waveguide case, we can only excite the emitters at the ENZ resonance, because this is a waveguide system that can couple energy inside its channel only around this particular resonance frequency. For the groove and rod waveguides, the emitters' excitation is not a problem, since we can excite the emitters at any frequency point from outside these plasmonic waveguides. Therefore, here, we set the pump frequency $\omega_p$ to be equal to the emission frequency $\omega_0$ of the emitters (corresponding wavelength $\lambda = 1018\text{nm}$), which is close to the ENZ resonance, leading to a detuning parameter equal to $\Delta_i = \omega_0 - \omega_p = 0$. Note, that the time scale (x-axis) is substantially elongated in Fig. 7 compared to the results in Fig. 5. Again, the ENZ waveguide has the highest concurrence values compared to the finite groove and rod waveguides for all types of pumping cases. In addition, the concurrence peaks of all waveguide systems are not very sensitive to the different pumping intensities. However, they rapidly decay when symmetric pumping is used (Fig. 7(b)).

Furthermore, we plot the contour maps of steady-state concurrence in Figs. 7(d)-(f) for a wide range of pumping intensities after a long time duration, $\gamma t = 70$, to better compare the entanglement efficiency mediated by the different plasmonic waveguides. It is clear that the largest steady-state concurrence is obtained by using the proposed ENZ plasmonic waveguide system, in contrast to the groove and rod cases, due to the relative large values of $\gamma_{12}$ and small values of $g_{12}$ that are independent of the emitters separation distance (Fig. 4, $d = 700\text{nm}$). Considering the different pumping strategies, we find that the antisymmetric pumping leads to much larger steady-state concurrence in the cases of the ENZ and groove waveguides, while the entanglement in the rod waveguide benefits from the symmetric pumping scheme (as it can be seen in Figs. 7(a)-(c)). It is also interesting to note that the pump strength should not be too large, in order to achieve strong steady-state entanglement between the qubits, otherwise strong interactions between the pump and qubits will occur and eventually lead to qubit decoupling and lasing [18]. Therefore, the contour plots presented in Figs. 7(d)-(f) demonstrate the lower and upper threshold values for the pump intensities in each passive waveguide case, that still preserve the entanglement between the qubits. The establishment of these limits is important for several quantum optical applications.

The panels in Fig. 8 demonstrate the steady-state concurrence after a long period of time ($\gamma t = 70$, similar to Fig. 7) as a function of the emitters' separation distance for the ENZ (left panels), finite groove (middle panels), and cylindrical rod (right panels) waveguides. Note, that for certain separation distances, a strong steady-state entanglement can be achieved by using all types of plasmonic waveguides when the pumping is relatively weak. However, it is further proved that the unique property of the ENZ waveguide is that the high steady-state concurrence does not depend on the inter-emitter separation distance [Fig. 8(a)]. On the contrary, in the cases of the finite groove and rod waveguide, the steady-state concurrence can be high or very low depending on the separation distances between the emitters. In addition, the strong and homogeneous entanglement is observed at the passive ENZ waveguide [Fig. 8(a)] only for the asymmetric or antisymmetric pumping and with relatively low pump intensities, consistent with the results presented in Fig. 7. These results further validate the importance of ENZ as an ideal nanophotonic environment to enhance entanglement-based applications.

Next, we calculate the second-order correlation function for the entangled steady-state, which is a metric that can be measured experimentally. In previous works, it was demonstrated that the degree of photon coherence given by the second-order photon correlation function is related to the concurrence that was used to characterize the entanglement [17,57]. This quantity can be calculated using the density matrix elements and for a zero time delay is equal to: [4]

$$g_{12}^{(2)} = \langle \rho \sigma_1^\dagger \sigma_2^\dagger \sigma_2 \sigma_1 \rangle / \langle \rho \sigma_1^\dagger \sigma_1 \rangle \langle \rho \sigma_2^\dagger \sigma_2 \rangle \tag{13}$$

It has been demonstrated that antibunching ($g_{12}^{(2)}(0) \to 0$) at zero time delay corresponds to high steady-state concurrence and, as a consequence, to strong qubit entanglement [17]. Therefore, measurements of the two-photon correlation function at zero delay can be used for entanglement detection.

To verify the relationship between $C_{ss}$ and $g_{12}^{(2)}(0)$, we plot in Fig. 9 the contour maps of the zero-delay second-order correlation function $g_{12}^{(2)}(0)$ for a wide range of pumping intensities. The emitters' separation distance is fixed to the same value used in Fig. 7 ($d = 700$ nm). As it can be seen by comparing Figs. 7(d)-(f) and 9(a)-(c), the antibunching signature coincides with the high steady state concurrence values. When $C_{ss}$ is large, the corresponding $g_{12}^{(2)}(0)$ is close to zero (antibunching), and when $C_{ss}$ goes to zero, $g_{12}^{(2)}(0)$ grows to large values, consistent with the results presented in [17] and [57] for groove and cylindrical rod (not ENZ) waveguides, respectively. In addition, the plot in Fig. 9 clearly shows that antibunching ($g_{12}^{(2)}(0) \to 0$) occurs for a much broader pumping range only in the case of the passive ENZ plasmonic system and not with the other plasmonic waveguides, confirming another practical advantage of the presented ENZ waveguide system.

Finally, we compute and plot in Fig. 10 the second-order correlation function at zero time delay, $g_{12}^{(2)}(0)$, as a function of the inter-emitter separation distance for the three plasmonic waveguides and for fixed pumping intensity values. Interestingly, the computed $g_{12}^{(2)}(0)$ value is almost constant for the ENZ waveguide case, when we spatially vary the relative positions of the two emitters, in agreement with the results presented in Fig. 8(a), where the steady-state entanglement $C_{ss}$ was found to be independent of the emitters' position and separation distances in the ENZ system. This constitutes the major advantage of the proposed ENZ plasmonic system when it comes to the experimental detection and measurement of the envisioned strong multi-qubit entanglement, especially in the common practical scenario of quantum emitters arbitrarily placed inside the resonating plasmonic system, since it is extremely difficult to accurate place emitters in nanoscale regions. On the contrary, the $C_{ss}$ strongly fluctuates as a function of the inter-emitter separation distance for the finite groove and rod waveguides, which means that the emitters need to be accurately placed at particular nanoscale spots along these nanowaveguides to detect the entanglement. Hence, a potential entanglement experiment based on ENZ waveguides is expected to be easier and more practical to be performed compared to similar experiments by using the other plasmonic waveguide cases.

**Conclusions**

We have demonstrated enhanced and prolonged in time long-range entanglement and giant RET enhancement between quantum emitters (qubits) placed inside ENZ metamaterial waveguides. The entanglement and RET are independent of the emitters' positions and separation distances due to the peculiar properties of the ENZ waveguides operating near their cutoff frequency, i.e., infinite phase velocity of the tunnelled electromagnetic wave and a phase that is uniform along the nanochannels. These interesting features, combined with the uniquely strong and homogeneous electromagnetic fields inside the ENZ waveguides, are ideal conditions to boost coherent light-matter interactions over long distances and increase the energy transfer rate between a donor-acceptor pair placed inside the nanochannels. In particular, the RET can be enhanced by at least two orders of magnitude by using the ENZ system with respect to the widely studied V-shaped and cylindrical rod plasmonic waveguides. Moreover, the transient and steady-state entanglement between quantum emitters can be improved and prolonged in time by placing them in arbitrary positions inside the ENZ waveguide nanochannel.

Since it is very difficult to control the position of emitters in nanoscale regions, the proposed ENZ plasmonic waveguide is advantageous for the experimental verification of the presented strongly entangled states compared to the alternative plasmonic waveguides (groove or rod), where the emitters need to be accurately placed in predefined positions to achieve maximum entanglement. This distinct feature makes the ENZ waveguide an ideal platform for multi-qubit entanglement operations. At the ENZ resonance, large spontaneous emission decay rate $\gamma$ and enhanced normalized decay rate $\gamma_{12}/\gamma$ can be satisfied simultaneously along the entire channel, consisting ideal conditions in order to achieve strong multi-qubit entanglement between two-level quantum emitters [33]. Moreover, the steady-state entangled state of the emitters placed inside the ENZ waveguide is robust enough to be observed experimentally by using second-order correlation function measurements because it persists over extended time periods and long distances. It is also demonstrated that the introduction of active (gain) media inside the plasmonic ENZ nanochannels can lead to perfect loss compensation of the inherent (nonradiative) plasmonic losses that, subsequently, reduces decoherence and further boosts transient quantum entanglement. We envision that the presented passive and active ENZ mediated RET and entangled states will find applications in future quantum information and communications integrated systems on a chip [58,59], the design of new low-threshold subwavelength nanolasers [60], and the creation of ultrasensitive quantum metrology devices [61]. Hence, effective ENZ waveguide media have the potential to become crucial components to the emerging field of quantum plasmonics [9].

## Acknowledgments

This work was partially supported by the National Science Foundation (DMR-1709612) and the Nebraska Materials Research Science and Engineering Center (MRSEC) (grant No. DMR-1420645).

**Figures**

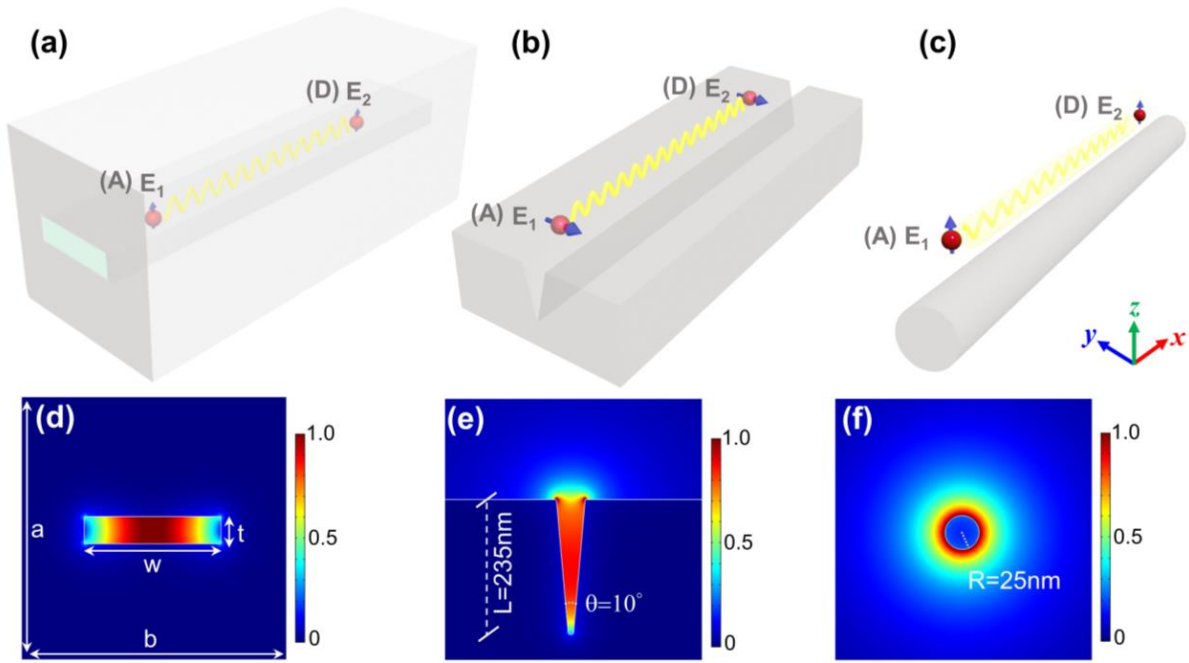

Figure 1 - Geometry of the proposed (a) ENZ plasmonic waveguide and two alternative plasmonic waveguides: (b) V-shaped groove and (c) cylindrical nanorod. All the presented waveguides are made of silver. A pair of two-level emitters [the donor (D) and acceptor (A)] are placed inside the nanochannel or along the different plasmonic waveguide structures. The lower panels (d)-(f) display the normalized fundamental mode distributions in the transverse cross section (yz-plane) at the same wavelength $\lambda = 1018$ nm and the geometry dimensions of each waveguide.

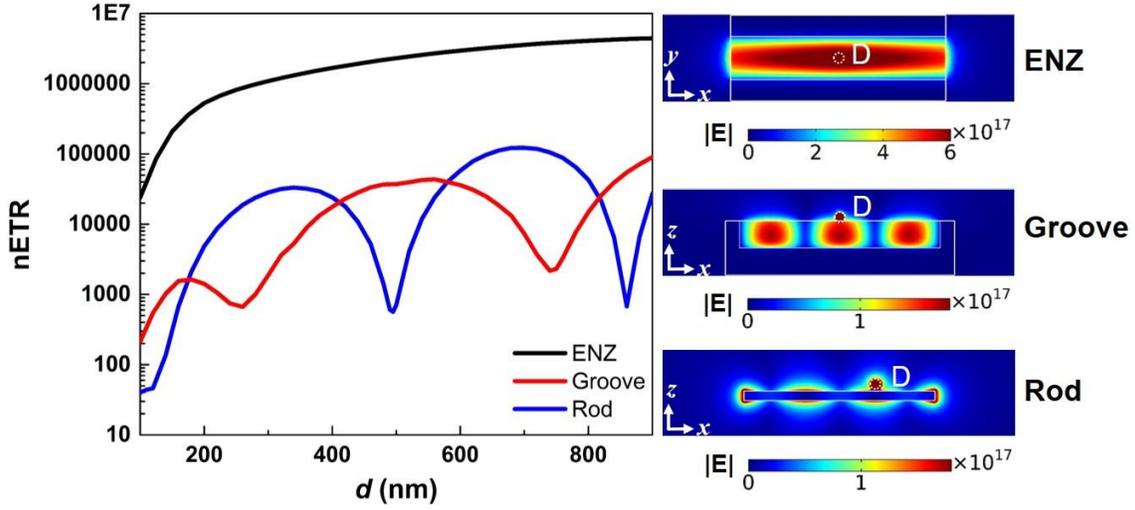

Figure 2 - The computed nETR values as a function of the separation distance d between the donor-acceptor pair in the cases of the ENZ, finite groove, and cylindrical rod plasmonic waveguides (see Fig. 1(a) for the specifications of the waveguides geometries). We assume the location of the acceptor to be fixed in the beginning of the waveguide channel while the donor position is being continuously moved along the x-axis. Right insets: The static electric field patterns when the donor with the electric dipole moment polarized along the z-axis is placed in the center (ENZ, groove) or slightly to the right off the center (rod) of each waveguide.

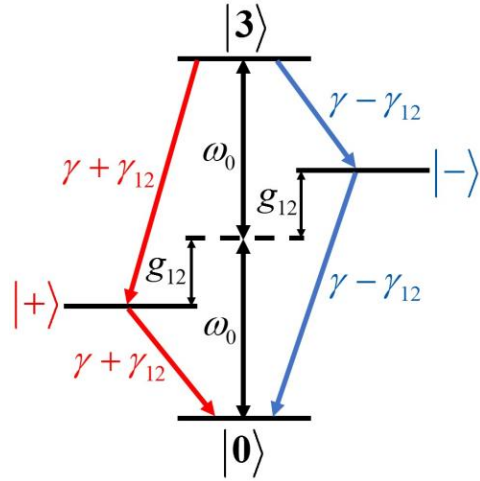

Figure 3 - Collective states diagram of two identical emitters. The two identical qubits (emitters) are located at equivalent positions with respect to the plasmonic system and have identical orientations.

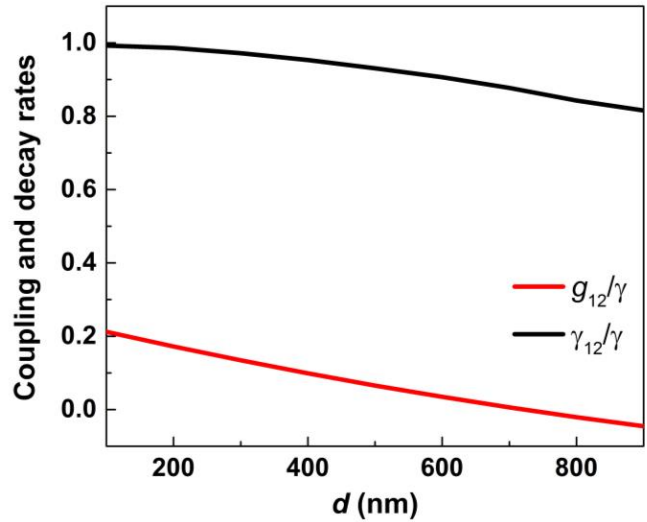

Figure 4 – Normalized decay rate $\gamma_{12}/\gamma$ and dipole-dipole interaction $g_{12}/\gamma$ of qubits placed inside the ENZ plasmonic waveguide as a function of the emitters' separation distance $d$. The obtained values are normalized to $\gamma$, which is the spontaneous emission decay rate of a single emitter placed in the waveguide.

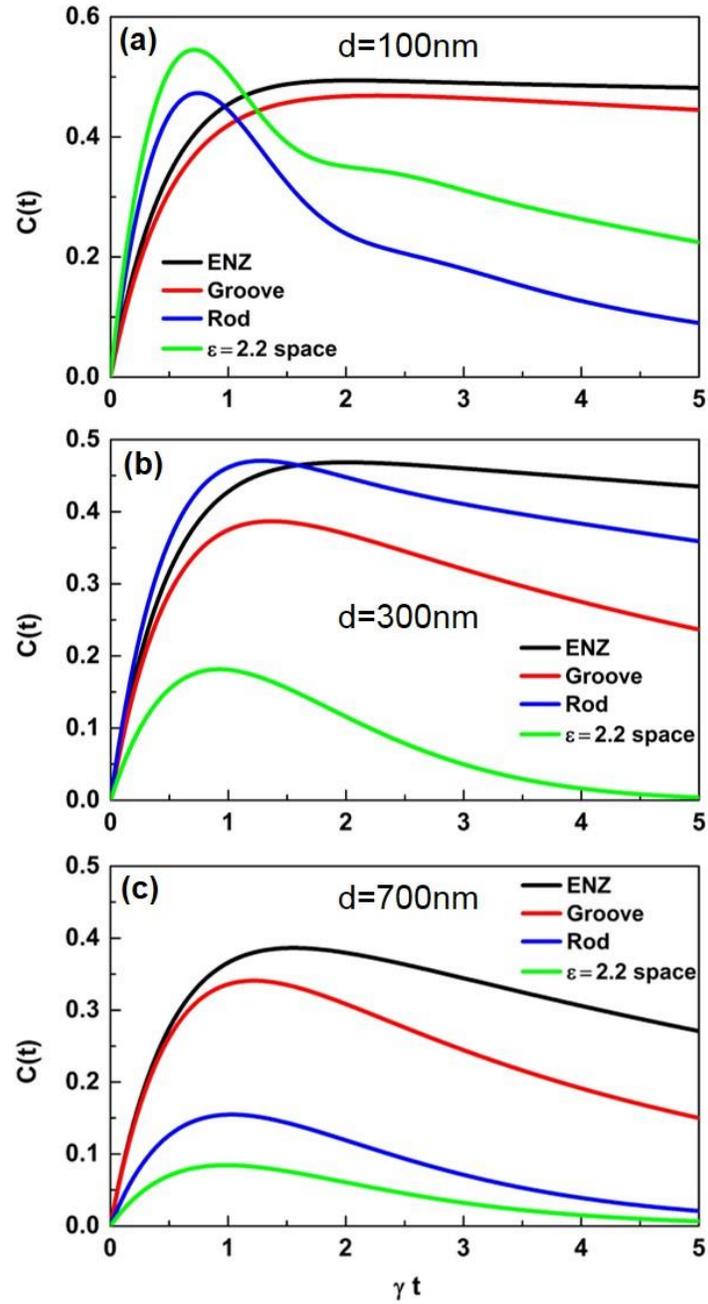

Figure 5 - Transient concurrence between the two qubits placed in the ENZ (black lines), groove (red lines), and cylindrical rod (blue lines) nanowaveguides for three inter-qubits separation distances: (a) $d = 100$ nm, (b) $d = 300$ nm, and (c) $d = 700$ nm. The green lines refer to the qubits placed in a dielectric bulk medium (no waveguide) with relative permittivity $\varepsilon = 2.2$.

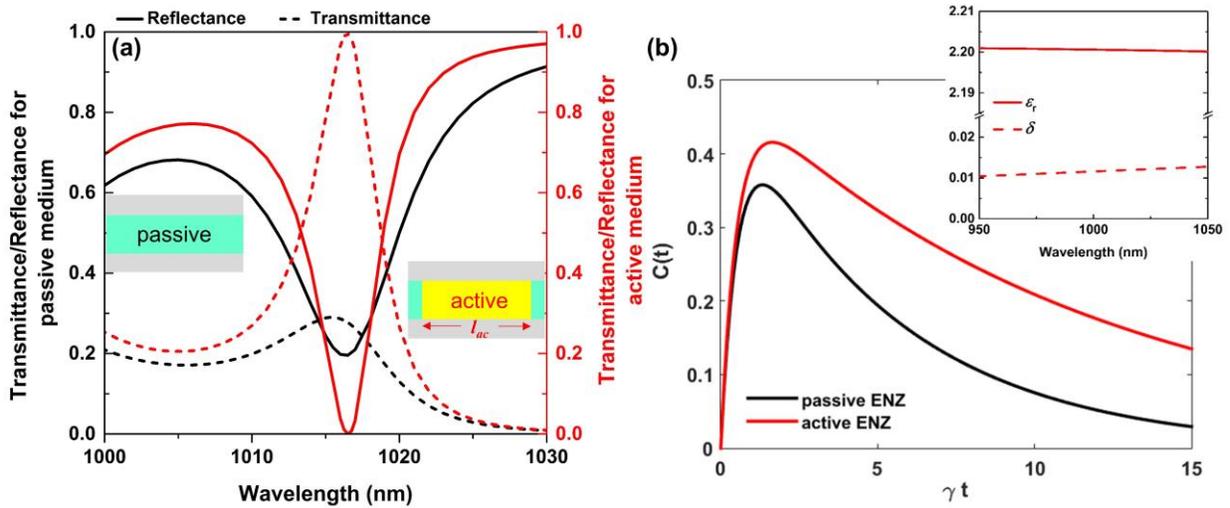

Figure 6 – (a) Transmittance and reflectance of the plasmonic waveguide in the vicinity of the ENZ resonance with (red) and without (black) gain. The left inset is the geometry of the passive ENZ plasmonic waveguide, and the right inset represents the active ENZ plasmonic waveguide. (b) Transient concurrence between the two qubits in the passive (black) and active (red) ENZ nanowaveguides for a fixed inter-qubits separation distance d = 900 nm. Inset: The Lorentz permittivity model of the realistic active dielectric material loaded inside the ENZ waveguide.

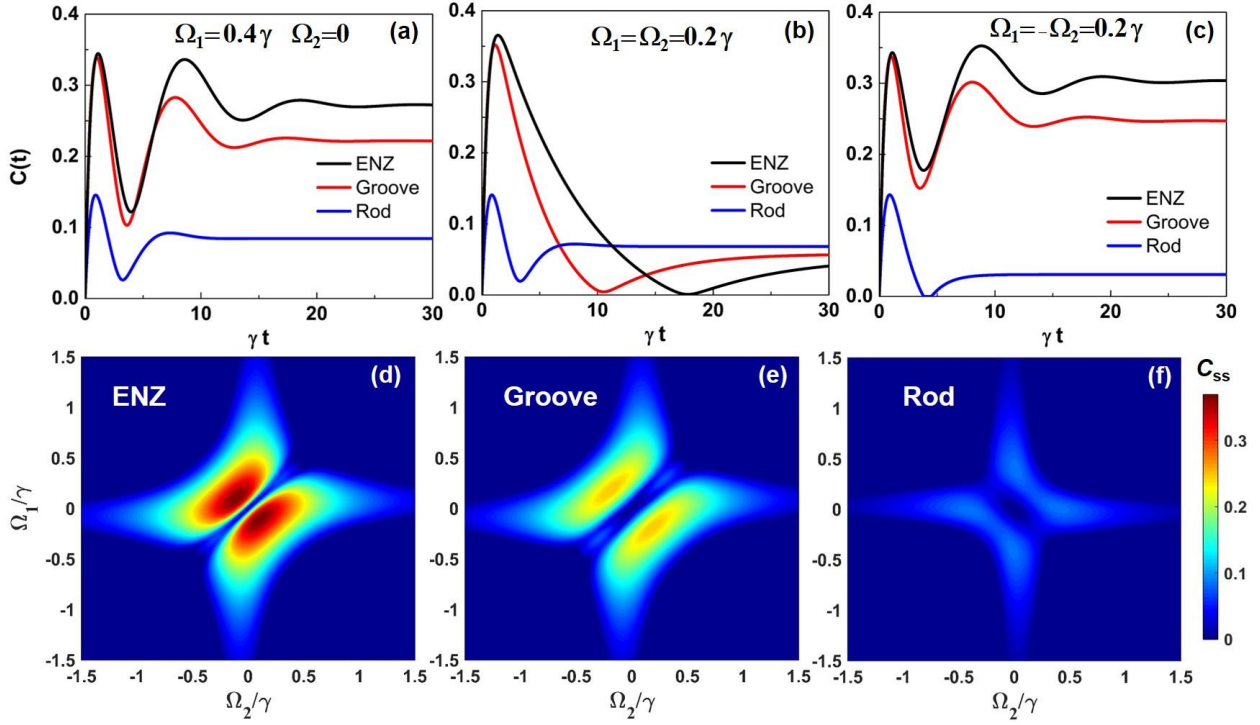

Figure 7 - Time dependence of the concurrence between two qubits pumped by (a) asymmetric, (b) symmetric, and (c) antisymmetric pumping. Lower panels (d)-(f): Steady state concurrence vs pumping intensities for the cases of (d) ENZ, (e) groove, and (f) cylindrical rod plasmonic waveguides. The emitters' separation distance is fixed to d =700nm in these results.

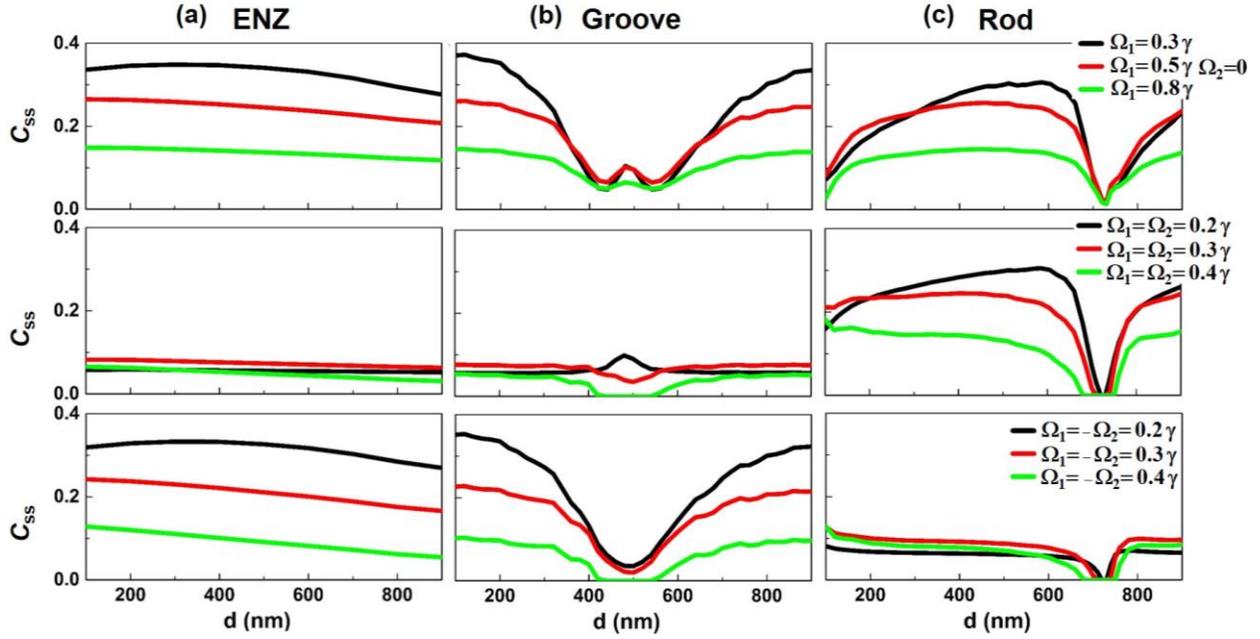

figure 8 - Steady state concurrence as a function of the qubits' separation distance monitored at the normalized time $\gamma t = 70$ for the cases of (a) ENZ, (b) groove, and (c) cylindrical rod plasmonic waveguides by using the asymmetric, symmetric, and antisymmetric pumping.

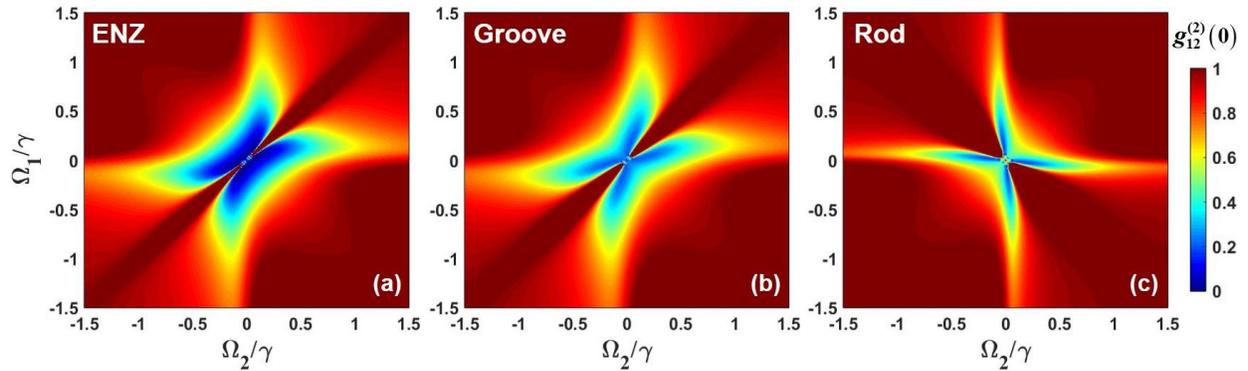

Figure 9 - Zero-delay correlations vs pumping intensities for the cases of (a) ENZ, (b) groove, and (c) cylindrical rod plasmonic waveguides, using the same parameters reported before in Figs. 7(d)-(f). The emitters' separation distance is fixed to d=700nm in these results.

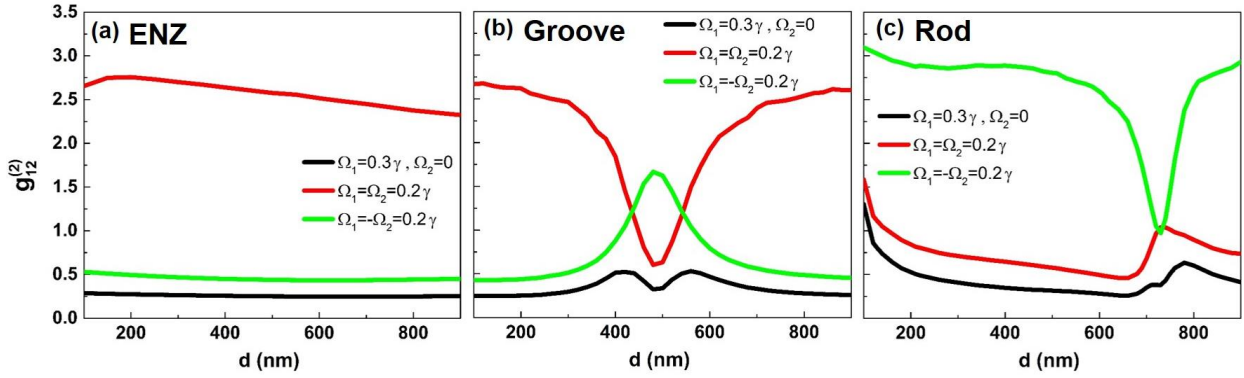

Figure 10 - A potential experimental validation of the qubit-qubit entanglement characterized by the second-order photon correlation function at zero time delay as a function of the emitters' separation distance for the cases of (a) ENZ, (b) groove, and (c) cylindrical rod plasmonic waveguides by using asymmetric, symmetric, and antisymmetric pumping.